\documentclass[12pt]{iopart}
\usepackage{epsf}
\begin{document}
\title[Cluster Transformation Coefficients ...]
{Cluster Transformation Coefficients for Structure and Dynamics
Calculations in {\it n}-Particle Systems:\newline
 Atoms, Nuclei, and Quarks}
\author{M Tomaselli\dag \footnote[3]{m.tomaselli@gsi.de}, L.C. Liu\ddag, T.
K\"uhl\dag, W. N\"ortersh\"auser\dag, D. Ursescu\dag,  and S.
Fritzsche\P}
\address{\dag\ GSI-Gesellschaft f\"ur Schwerionenforschung, D-64291 Darmstadt,
Germay.}
\address{\ddag\ T-Division,
Los Alamos National Laboratory, Los Alamos, NM 87545,USA.}
\address{\P\ Institute of Physics, Kassel University,
D-34132 Kassel, Germany.}
\begin{abstract}
The structure and dynamics of an {\it n}-particle system are
described with coupled nonlinear Heisenberg's commutator
equations where the nonlinear terms are generated by the two-body
interaction that excites the reference vacuum via
particle-particle and particle-hole excitations. Nonperturbative
solutions of the system are obtained with the use of dynamic
linearization approximation and cluster transformation
coefficients. The dynamic linearization approximation converts
the commutator chain into an eigenvalue problem. The cluster
coefficients factorize the matrix elements of the {\it
(n)}-particles or particle-hole systems in terms of the matrix
elements of the {\it (n-1)}-systems coupled to a
particle-particle, particle-hole, and hole-hole boson. Group
properties of the particle-particle, particle-hole, and hole-hole
permutation groups simplify the calculation of these coefficients.
The particle-particle vacuum-excitations generate superconductive
diagrams in the dynamics of {\it 3}-quarks systems. Applications
of the model to fermionic and bosonic systems are discussed.
\end{abstract}
\maketitle
\section{Introduction}
In the Heisenberg's picture the time evolution of a system of
particles is described by a commutator equation which involves the
{\it n}-body Hamiltonian and the creation operators of the {\it
n}-body ground- and excited-modes. However, the excitations of
the reference vacuum, resulting from the scattering of particles
from the vacuum to higher states (the particle-hole and
particle-particle excitations), are completely neglected in this
formulation, namely, the time evolution of the {\it n}-body-modes
is described by a linearized Equation of Motion (EOM) [1] which
involves only valence particles. Recently the original Heisenberg
formulation of the {\it n}-body-dynamics has been generalized
within the framework of the Dynamic-Correlation Model (DCM)~[2]
for fermions, the Boson Dynamic-Correlation Model (BDCM)~[3] for
bosons, and the Superconductive Dynamic- Correlation Model (SDCM)
which describes superconductive- and polarization-effects~[4]. In
these models, the inclusion of the structure and the dynamics of
odd/even quasi-particles into the calculations of the excitations
of the model-vacuum led to the modification of the formal
strcture of the original Heisenberg's commutator equation. The
new dynamics system is characterized by a set of coupled
commutator equations which involve simultaneously the excitations
of the valence quasi-particles and the excitations of the
Intrinsic-Supercunductive-Vacuum States (ISVSs) {\it i.e.}:
mixed-mode states formed by coupling the valence quasi-particles
to the excitations of the vacuum (particle-particle and
particle-hole). As in  Ref.~[4], the resulting mixed-mode states
are classified in terms of the following mixed-mode wave
functions: (a) valence particles coupled to ISVSs formed by
include particle-hole, particle-particle and hole-hole
vacuum-excitation modes that have the same parity of the valence
particles; (b) valence particles and ISVSs formed by exciting
particle-hole, particle-particle and hole-hole having a parity
different from that of the valence particles. The superconductive
vacuum states (b) are not considered by perturbative theories,
although they may be associated to the creation of virtual
particles~[2,~3] giving important contribution to dynamic
theories. The superconductive vacuum excitation modes are
important mainly at high densities~[4], although the overlap of
the particle-particle excitation modes may influence, due to the
strong Pauli blocking effects in the ISVSs, the particle-hole
excitation mechanism also at medium energies. In this paper we
disregard the superconductive effects, which will be the subject
of a paper in preparation, and discuss only the effect of the
particle-hole excitation mechanism on ground state properties of
atomic, nuclear, and quark systems. In principle, the system of
commutator equations may be solved perturbatively by means of
substitution method which consists in inserting the higher order
commutators into the previous calculated commutators. However,
such perturbative method is usually not convergent for systems of
strong interacting particles and, therefore, will not be
discussed here. We will study nonperturbative solutions based on
the following: (a) a linearization ansatz motivated by the
consideration that in the low energy domain only few particle
excitations are energetically representative while the higher
excitation modes (with admixture probability less than 0.1 \%)
may  be omitted from the model space; (b) cluster transformation
coefficients which provide an exact solution for
 the complex {\it n}-body matrix elements that are input of the commutator
equations. The wave functions, solutions of the calculated
collective eigenvalue equation are classified as in Refs.~[2,~3]
in terms of Configuration Mixing Wave Functions (CMWFs). In
Section II, we calculate the commutator chain for one particle-,
two particle-, and three particle- creation operators. The
generalization of the model to hole- and (particle-hole)-valence
operators is not given explicitly because of easy extrapolation
from the present particle formulation. The commutator chain is
then linearized and converted into an eigenvalue problem which is
solved calculating the {\it n}-body matrix elements within the
cluster transformation coefficients. In Section III, solutions of
the nonlinear eigenvalue equation for one valence particle/hole
are obtained and the HFS constants of hydrogen- and lithium-like
heavy ions are calculated. In Section IV, the commutator chain is
solved for two and three valence particles
 and the resulting CMWFs are used to calculate the matter distribution of
halo nuclei. In Section V, an application to the three {\it
dressed} quark systems is discussed in order to provide new
theoretical data for the polarizability of the proton. For a
measurement of the proton polarization, novel ultra/intense
lasers may be useful, like the PHELIX-petawatt laser presently
under construction at GSI, have to be used. A direct way for a
precision measurement seems to be in reach by the latest
improvements of energetic multi-MeV photon sources using laser
backscattering at electron storage rings.

\section{The nonlinear commutator chain}
The nonlinear commutator chain is an extension of the
Heisenberg's equation. It allows us to address the situation where
valence clusters and core clusters are almost energetically
degenerated and may, therefore, coexist. We introduce valence
systems formed of either neutrons or protons. and define the
valence states as
\begin{eqnarray}
\label{equ-ii-1} \: \:|\Gamma_{J} (\{n\}~particles) \rangle \: =
\sum_{\alpha_n } \: X^{(n)}_{\alpha_n J_n} \: N^{(n)}_{\alpha_n
J_n} \: A^{\dagger}_n (\alpha_n J_n; J)|0  \rangle,
\end{eqnarray}
\vspace*{0.3cm} where $n=1,2,3$ and  $J$ is the total spin. The
$N^{(n)}_{\alpha_n J_n}$ are the normalization constants,
$X^{(n)}_{\alpha_n J_n}$ the mode amplitudes, and  $\alpha_1$
denotes the other quantum numbers of the valence particles. The
creation operator is defined by
\begin{eqnarray}
\label{equ-ii-2} A^{\dagger}_1(\alpha_1 j_1; J) \equiv
a^\dagger_{j_1}\,
\end{eqnarray}
for one valence particle (fermion);
\begin{eqnarray}
\label{equ-ii-3}
 A^{\dagger}_2(\alpha_2 J; J) \equiv (a^\dagger_{j_1} \otimes
a^\dagger_{j_2})^{J}\,
\end{eqnarray}
for one valence pair (boson);
\begin{eqnarray}
\label{equ-ii-4}
 A^{\dagger}_3(\alpha_3 J_1; J)\equiv ((a^\dagger_{j_1}\otimes
a^\dagger_{j_2})^{J_1}
 \otimes a^\dagger_{j_3})^{J}
\end{eqnarray}
for three valence particles (fermion). Hence, in
Eq.~(\ref{equ-ii-1}), $J_n=j_1$ for one particle, $J_n=J$ for two
particles, and $J_n=J_1$ for three particles.\newline In the
literature, various approaches, as reported in Ref.~[3], have been
proposed for including core excitations. In Shell-Model
calculations the residual interaction between the valence
particles excites the valence pairs to higher single particle
states, leaving the vacuum in the $J=0$ ground state. In Ref.~[5]
the core excitations have been included in the Shell-Model
calculations through introducing the coupling between the valence
particles and the collective $J=2^+$  core-excited states. In the
DCM and BDCM the core excitation is included through coupling the
valence fermionic/bosonic states,
Eqs.~(\ref{equ-ii-2},~\ref{equ-ii-3}) to intrinsic bosonic states
corresponding to particle-hole excitations of the nuclear core.
In this paper, we consider the following mixed-mode
fermionic/bosonic states:
 a)  valence fermionic/bosonic states coupled to
the dynamic particle-hole states of normal parity;
 b)  valence fermionic/bosonic states coupled to
the dynamic particle-hole states of non-normal parity. The
particle-hole coupling is implemented through the two-body force
$V$, such that
\begin{eqnarray}
\label{equ-ii-5} H = \sum_{\alpha} \: \epsilon_{\alpha} \:
a^{\dagger}_{\alpha} \: a_{\alpha} + \frac{1}{2} \: \sum_{\alpha
\beta \gamma \delta} \: v_{\alpha \beta \gamma \delta} \:
a^{\dagger}_{\alpha} \: a^{\dagger}_{\beta} \: a_{\delta} \:
a_{\gamma} = H_0 + V,
\end{eqnarray}
\vspace*{0.2cm} where $v_{\alpha \beta \gamma \delta}$ are the
matrix elements of the realistic two-body potential which has two
parts: a central part and a tensor part. The tensor component of
the realistic two-body potential shapes the many-body Hamiltonian
and in particular its long tail acting between the valence
fermions/bosons and causing simultaneously the excitation of the
particles to high shell-model states and the deformation of the
nuclear core.

The mixed-mode $\{1-2-3\}~particles \{1'-2'\}~bosons$ states,
denoted $\mid \Gamma^{\prime}_J \rangle,$ can therefore be
expanded as follows:
\begin{eqnarray}
\label{equ-ii-6} \fl{ |\Gamma^{\prime}_{J}
(\{n=1-2-3~particles\}-\{1'-2'\}~bosons) \rangle } ={ \left [
\sum_{\alpha_n J_n} X^{(n)}_{\alpha_n J_n} N^{(n)}_{\alpha_n; J_n}
A^{\dagger}_n (\alpha_n J_n; J)  \right.  } \nonumber\\
\fl+{ \sum_{\alpha_{n+1'}J_n J_{n+1'}}
X^{(n+1')}_{\alpha_{n+1'}J_n J_{n+1'}}
N^{(n+1')}_{\alpha_{n+1'}J_n J_{n+1'}}
A^{\dagger}_{n+1'} (\alpha_{n+1'}(J_n J_{n+1'}; J))}\nonumber \\
\fl+{\left. \sum_{\alpha_{n+2'}J_n J_{n+1'} J_{n+2'}}
X^{(n+2')}_{\alpha_{n+2'} J_n J_{n+1'} J_{n+2'}}
N^{(n+2')}_{\alpha_{n+2'} J_n J_{n+1'} J_{n+2'}}\right.} {\left .
A^{\dagger}_{n+2'} (\alpha_{n+2'}(J_n J_{n+1'} J_{n+2'};
J))\right]|0 \rangle},
\end{eqnarray}
 where $1, 2, 3$ labels the particles and $1', 2'$ label the bosons
(particle-hole pairs). The operators
\begin{eqnarray}
\label{equ-ii-7} \fl { A^{\dagger}_{n+1'} (\alpha_{n+1'} (J_1
J_{n+1'}); J) \equiv ([ A^{\dagger}_n (\alpha_n (J_n; J) \otimes
(a^\dagger_{j_{n+1}} \otimes
b^{\dagger}_{j_{n+2}})^{J_{n+1'}}]^{J_{(n,n+1')}})^J,}
\end{eqnarray}
\begin{eqnarray}
\label{equ-ii-8} \fl { A^{\dagger}_{n+2'}(\alpha_{n+2'}(J_n
J_{n+1'} J_{n+2'}); J)
\equiv ( [ A^{\dagger}_{n}(\alpha_{n}(J_n; J)) \otimes } \nonumber\\
\fl { ((a^\dagger_{j_{2n+1}} \otimes
b^{\dagger}_{j_{2n+2}})^{J_{n+1'}} \otimes(a^\dagger_{j_{2n+3}}
\otimes b^{\dagger}_{j_{2n+4}})^{J_{n+2'}})
     ^{J_{(n+1',n+2')}}]^{J_n,J_{(n+1',n+2')}} )^J}
\end{eqnarray}

contain consequently the hole creators $b^{\dagger}_j$. To
conclude, in the DCM and BDCM one starts with the valence
fermions/bosonic states of Eqs.~(\ref{equ-ii-2},~\ref{equ-ii-4})
and constructs subsequently mixed-mode nuclear states by including
components having additional bosons formed by the particle-hole
pairs of the core excitations. The resulting nuclear states are
then classified in terms of (CMWFs) of increasing degrees of
complexity (number of particle-hole pairs) Ref.~[2]. The basic
dynamic equations of the models are the commutator equations that
involve the nuclear many-body Hamiltonian $H$ and the operators
$A^{\dagger}_n (n=1, 2, 3)$. After a lengthy but elementary
algebra, we obtain the following results:

(a) Commutator equation for $\{n=1, 2, 3\}$ states:
\begin{eqnarray}
\fl{[H, A^{\dagger}_n (\alpha_n J_n; J)] } ={ \sum_{\beta_n J'_n}
\: \langle A_n(\alpha_n J_n; J) \| H \| A^{\dagger}_n (\beta_n
J'_n; J) \rangle \: }
 {A^{\dagger}_n (\beta_n J'_n; J) }\nonumber\\
\fl+{ \sum_{\beta_{n+1'} J'_{n} J'_{n+1'}} \: \langle A_n
(\alpha_n J_n; J)\|H\| A^{\dagger}_{n+1'} (\beta_{n+1'} J'_n
 J'_{n+1'}; J)  \rangle \: }\nonumber\\
\fl \times {  A^{\dagger}_{n+1'} (\beta_{n+1} J'_n J'_{n+1'}; J).}
\label{equ:II-10}
\end{eqnarray}
(b) and Commutator equation for $\{n=1, 2, 3; n'=1'\}$  states
\begin{eqnarray}
\fl{[H, A^{\dagger}_{n+1'} (\alpha_{n+1'} J_n J_{n+1'}; J)] } \nonumber\\
\fl ={ \sum_{\beta_{n+1'} J'_n J'_{n+1'}} \: \langle A_{n+1'}
(\alpha_{n+1'} J'_n J_{n+1'}; J) \| H \| A^{\dagger}_{n+1'}
(\beta_{n+1'} J'_n J'_{n+1'}; J)
\rangle \:}\nonumber \\
\fl \times {A^{\dagger}_{n+1'} (\beta_{n+1'} J'_n J'_{n+1'}; J) }\nonumber\\
\fl+{ \sum_{\beta_{n+2'} J'_n J'_{n+1'} J'_{n+2'})} \: \langle
A_{n+1'} (\alpha_{n+1'} J_n J_{n+1'}; J)\|H\| A^{\dagger}_{n+2'}
(\beta_{n+2'} J'_n
J'_{n+1'} J'_{n+2'}); J)  \rangle \: }\nonumber\\
\fl \times {  A^{\dagger}_{n+2'} (\beta_{n+2'} J'_n J'_{n+1'}
J'_{n+2'}; J).} \label{equ:II-11}
\end{eqnarray}

The commutator equations are used to obtain the solutions of DCM
with use of the Equations of Motion (EOM) method~[6]. In this
latter method, one looks for an operator $C^{\dagger}_{m}$ such
that $C^{\dagger}_{m}|0\rangle = |m\rangle $ with $H|m\rangle =
E_{m}|m\rangle$ and $H|0\rangle = E_{0}|0\rangle$ $(E_{0} \equiv
0)$. Here $|m\rangle$ denotes the excited state and $|0\rangle$
the correlated ground state. One then has the operator identity
$[H, C^{\dagger}_{m}] = E_{m}C^{\dagger}_{m} $. Upon replacing
$C^{\dagger}$ by $A^{\dagger}$ and using the results in
Eqs.~(\ref{equ:II-10},~\ref{equ:II-11}) for the l.h.s. of the
above operator identity, one obtains a set of equations which,
after linearization, can be transformed into a system of
eigenvalue equations. As one can see, the linearization consists
in approximating the higher-order diagrams by an effective term.
The solutions of the linearized commutator equations, can,
therefore, be regarded as eigenvalues of a model fermionic/bosonic
Hamiltonian.

The valence particles become, therefore, the {\it dressed}
solutions of the following collective nonlinear Hamiltonian~[7]:
\begin{eqnarray}
H_{coll} = (T_{coll} + V_{coll})= \sum_i^N \: E_{coll}^i \: {\cal
A}^{\dagger}_i (\alpha_i J_1 \ldots J_i; J) \: {\cal A}_i
(\alpha_i J_1 \ldots J_i; J) \label{equ-ii-12}
\end{eqnarray}
with
\begin{eqnarray}
\fl{{\cal A}^{\dagger}_i (\alpha_i (J_1 \ldots J_i); J) =
\left[X^{i}_{\alpha_i J_i \ldots J_i} A^{\dagger}_i (\alpha_i J_i
\ldots
J_i; J)\right.}\nonumber\\
\fl+{\left. \: X_{\alpha_{i+1'} J_i  \ldots J_{i+1'}}
A^{\dagger}_{i+1'} (\alpha_{i+1'} J_1 \ldots J_{i+1'}; J)
 + X_{\alpha_{i+2'} J_i  \ldots J_{i+2'}}
A^{\dagger}_{i+2'} (\alpha_{i+2'} J_i \ldots J_{i+2'}; J) \right]
}
 \end{eqnarray}
\vspace{.2cm} where the $X_{\alpha}$ are the calculated
mixed-mode amplitudes.

The input to solving the commutator equations are the matrix
elements of the $n-particle$ configuration mixing wave functions
(CMWFs). The latter can be easily calculated by use of the
cluster-factorization method of Ref.~[2,~3]. The method is guided
by the observation that the EOM connect the $n-$particle states
to the $(n-1)$-particle states. In the DCM, the $n\rightarrow
n-1$ reduction is achieved by factorizing out 1 boson, which can
be $\{2p\}$, $\{1p-1h\}$, or $\{2h\}$. In the following, we
exemplify the method by considering the $\{4p-2h\}$ $(n=3)$
parent configuration. For the sake of simpler notation, we will
not write the detailed coupling of the creation operators but
write the wave functions $\Psi$. There are three active pairs
associated with the $\Phi^3_{JM} (\alpha_3 J_1 J_2 J_3)$
configuration. To go to the $(n=2)$ (two active pairs)
configuration, we can factorize out either $\{1p-1h\}$, $\{2p\}$,
or $\{2h\}$ to arrive at $\{3p-1h\}$, $\{2p-2h\}$, or $\{4p\}$,
respectively. This leads to the following expansion in terms of
$\{3p-1h\}$-CMWFs with coordinates $\{\alpha_2 \}$,
$\{2p-2h\}$-CMWFs with coordinates $\{ \lambda_2 \}$, and
$\{4p\}$-CMWFs with coordinates $\{ \epsilon_2 \}$:
\begin{eqnarray}
\fl{|\Phi^{4p2h}_{J} (\alpha_3 J_1 J_2 J_3) \rangle \: =}\nonumber\\
\fl={ \frac{1}{\sqrt{3}} \sum_{\alpha_2 \overline{\alpha}_2
J_{k_1} J_{k_2} J_r J_s} \: ^3T_J (\alpha_3 J_1 J_2 J_3 | \}
\alpha_2 J_r \overline{\alpha}_2 J_s) \: \left[|\Psi^{3p1h}_{J_r}
(\alpha_2 J_{k_1} J_{k_2})\rangle \: \otimes |\Psi^{1p1h}_{J_s}
(\overline{\alpha}_2 J_s)\rangle \right]^J }\nonumber\\
\fl+{ \frac{1}{\sqrt{3}} \sum_{\lambda_2 \overline{\lambda}_2
J_{k_1} J_{k_2} J_r J_s} \: ^3V_J (\alpha_3 J_1 J_2 J_3| \}
\lambda_2 J_r \overline{\lambda}_2 J_s) \left[ |\Psi^{2p2h}_{J_r}
(\lambda_2 J_{k_1} J_{k_2})\rangle \: \otimes |\Psi^{2p}_{J_s}
(\overline{\lambda}_2 J_s)\rangle \right]^J }\nonumber\\
\fl+{ \frac{1}{\sqrt{3}} \sum_{\epsilon_2 \overline{\epsilon}_2
J_{k_1} J_{k_2} J_r J_s} \: ^3Z_J (\alpha_3 J_1 J_2 J_3 | \}
\epsilon_2 J_r \overline{\epsilon}_2 J_s) \left[ |\Psi^{4p}_{J_r}
(\epsilon_2 J_{k_1} J_{k_2})\rangle \: \otimes |\Psi^{2h}_{J_s}
(\overline{\epsilon}_2 J_s)\rangle \right]^J \: \: \: .}
\label{equ-ii-13}
\end{eqnarray}

\begin{figure}
\epsfbox{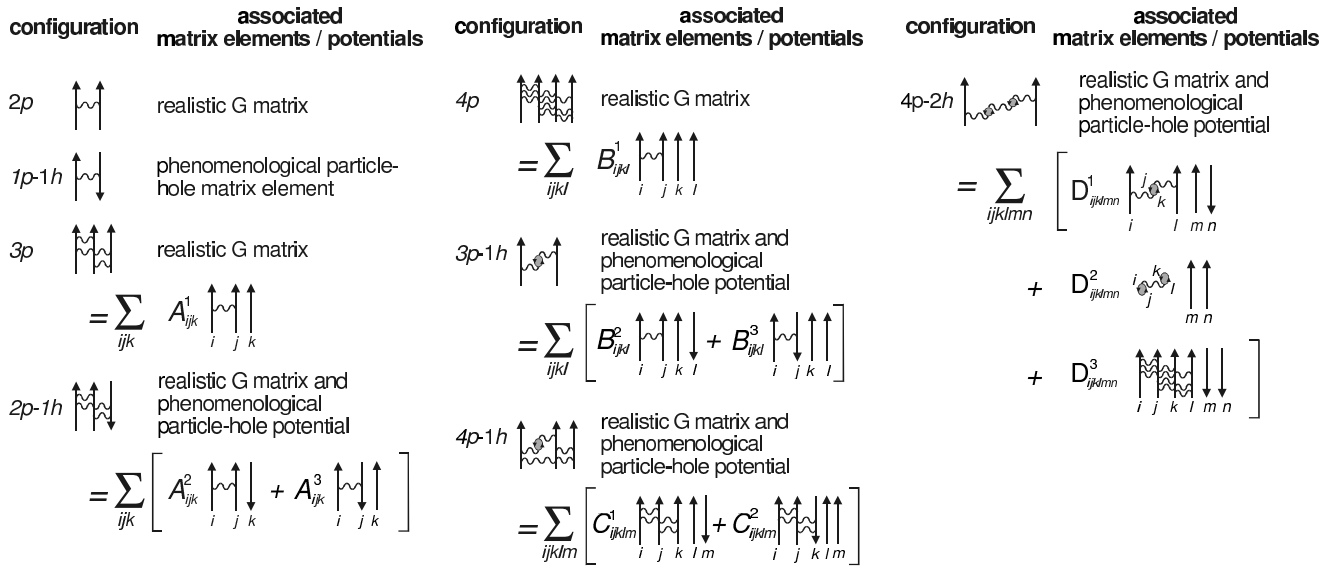} \vspace{-18cm}
\caption{\label{ME} Factorisation of the matrix elements in terms
of cluster coefficients.}
\end{figure}

\vspace*{0.3cm} \noindent The cluster transformation coefficients
in Eq.~(\ref{equ-ii-13}), i.e: the
 $^3T_J (\alpha_3
J_1 J_2 J_3 | \} \alpha_2 J_r \overline{\alpha}_2 J_s)$, the
$^3V_J (\alpha_3 J_1 J_2 J_3 | \} \lambda_2 J_r
\overline{\lambda}_2 J_s)$, and the $^3Z_J (\alpha_3 J_1 J_2 J_3
| \} \epsilon_2 J_r \overline{\epsilon}_1 J_s)$ are then
calculated, as in Ref.~[2,~3], by reducing the $SU_{2J+1} (3)$
representations carried by the wave functions on the right-hand
side of Eq.~(\ref{equ-ii-7}), {\it i.e.}, by diagonalizing the
Casimir operators [8] of the $SU_{2J+1} (3)$ groups in the basis
states of Eq.~(\ref{equ-ii-13}). Hence, the coefficients $^3T_J
(\alpha_3 J_1 J_2 J_3 | \} \alpha_2 J_r \overline{\alpha}_2 J_s)$
are eigenvalues of the following  matrix:
\begin{eqnarray}
\fl{\left( ^3T_J (\alpha_3 J_1 J_2 J_3| \} \alpha_2 J_r
\overline{\alpha}_2 J_s) \right)^{\dagger} \: ^3T_J (\alpha_3 J_1
J_2 J_3 |
\} \alpha_2 J_r \overline{\alpha}_2 J_s) =}\nonumber\\
\fl={ \sum_{k J_i J'_i J^2_r J^3_r} \: (-1)^{J_i + J'_i + J^2_r +
J^3_r + J_s + J'_s} \: (\hat{k})^{1/2} \: \hat{J}_r \hat{J}'_r \:
}\nonumber\\
\fl{ \: \left\{ \begin{array}{ccc} J_r & J & J_s \\ J & J_r & k
\end{array} \right\} \: \left\{ \begin{array}{ccc} J'_r & J & J'_s \\ J & J'_r
& k \end{array} \right\} \: \left\{ \begin{array}{ccc} J_i & J_r & J^2_r \\
J_r & J_i & k \end{array} \right\} \: \left\{ \begin{array}{ccc}
J'_i & J'_r &
J^3_r \\ J'_r & J'_i & k \end{array} \right\} \: }\nonumber\\
\fl{ \: \left( ^2 T^k_{J_r} (\alpha_2 J_{k_1} J_{k_2} | \}
\alpha_1 J_i \overline{\alpha}_1 J^2_r) \right)^{\dagger} \:
^2T^k_{J_r} (\alpha_2 J_{k_1} J_{k_2}| \}\alpha_1 J_i
\overline{\alpha}_1 J^2_r) \:
}\nonumber\\
\fl{ \: \left(^2T^k_{J'_r} (\beta_2 J_{k_3} J_{k_4} | \} \beta_1
J'_i \overline{\beta}_1 J^3_r) \right)^{\dagger} \: ^2T^k_{J'_r}
(\beta_2 J_{k_3} J_{k_4} | \} \beta_1 J'_i \overline{\beta}_1
J^3_r) \: \: \: .} \label{equ-ii-14}
\end{eqnarray}
In Eq.~(\ref{equ-ii-14}), the sum over $J_{k_1}$, $J_{k_2}$,
$J_{k_3}$, $J_{k_4}$, $\alpha_1$, $\overline{\alpha_1}$,
$\beta_1$, and $\overline{\beta_1}$ has not been written
explicitly but is understood. The 6$J$ coefficients are defined
according to Ref.~[9]. The same procedure has been applied to
calculate the $^3V$ and $^3Z$ cluster-transformation coefficients.
The cluster-transformation coefficients of the $^{4}T$ are then
expanded in terms of $^3T$ using Eq.~(\ref{equ-ii-14}) in the
limit $^{4}T \rightarrow ^{3}T$ and $^{3}T \rightarrow ^{2}T$. In
tables~\ref{3p} to \ref{4p1h2} the transformation coefficients are
exemplified.
 The above recursive procedure is equally valid for
cluster transformation coefficients involving fermionic CMWFs.
Within the cluster transformation coefficients introduced for the
fermionic/bosonic CMWFs, the matrix elements needed for the EOM
can be readily obtained:

The matrix elements for two particles are
\begin{eqnarray}
\fl{ (a) \langle \Phi^{2p}_J |V| \Phi^{2p}_J \rangle =
\langle \Phi^{2p}_J |G(E)| \Phi^{2p}_J \rangle }\nonumber\\[10pt]
\fl{ (b)\langle \Psi^{1p1h}_J| V | \Psi^{1p1h}_J \rangle =\langle
\Psi^{1p1h}_J| V_{phen} | \Psi^{1p1h}_J \rangle }
\label{equ-ii-63}
\end{eqnarray}
In Eq.~(\ref{equ-ii-63}), G(E) is the realistic G-matrix, and
$V_{phen}$ is a phenomenological potential. The matrix elements
for:
\begin{eqnarray}
\fl{ (c) \{3p\}, \{2p-1h\}, \{2h-1h\}, \{3p-1h\}, \{4p\},
\{4p-1h\}, \{4p-2h\} }
\end{eqnarray}
 are given symbolically in Fig.~\ref{ME}.
\section{HFS in hydrogen- and lithium-like ions}
With the reduction of the  CMWFs and the factorization of the
{\it n}-body matrix elements at our disposal we have performed
calculations for the hyperfine splitting (HFS) of heavy ions~[10].
In table~\ref{tab-DCM-QED} the DCM results are compared to
ground-state HFS splittings calculated for a point nucleus and
with results which take the finite spatial distribution of the
nuclear charge into account (Breit-Schawlow correction).
Additionally, QED radiative corrections~[15] are listed and
combined with the DCM results. The pure DCM splittings agree
remarkably well with the experimental values, which are given in
the last row of the table, while the wavelengths obtained after
adding the QED contributions show a systematic shift to larger
wavelengths. To clarify this still open point and in order to
obtain further informations about the hyperfine interaction of
the nucleus with the electronic cloud of high-$Z$ ions , we have
also performed calculations for the HFS of the lithium-like ions.
The result of the calculation~[16] of the HFS for
$^{209}$Bi$^{80}$ is given and compared with other theories in
table~\ref{lilike}. In table~\ref{lilike} the experimental result
of Ref.~[19] is also given. A new experiment will be performed at
GSI (2003) in order to localize the resonance within an even
smaller error. The {\it boiling} of the QED vacuum terms given in
table~\ref{lilike}, i.e. the electron-positrons contributions up
to now not considered by other theories, may of course also
generate a double counting effect with the QED contributions in
lithium-like ions~[16]. In the DCM the mixed-mode (b) states
generate contributions to the HFS that may be responsible for a
double counting effect with the QED perturbative calculation. The
terms (b) are not considered by any perturbative calculation
performed for the magnetic structure of nuclei.
\section{Elastic proton scattering on exotic nuclei}
\begin{figure}
\epsfbox{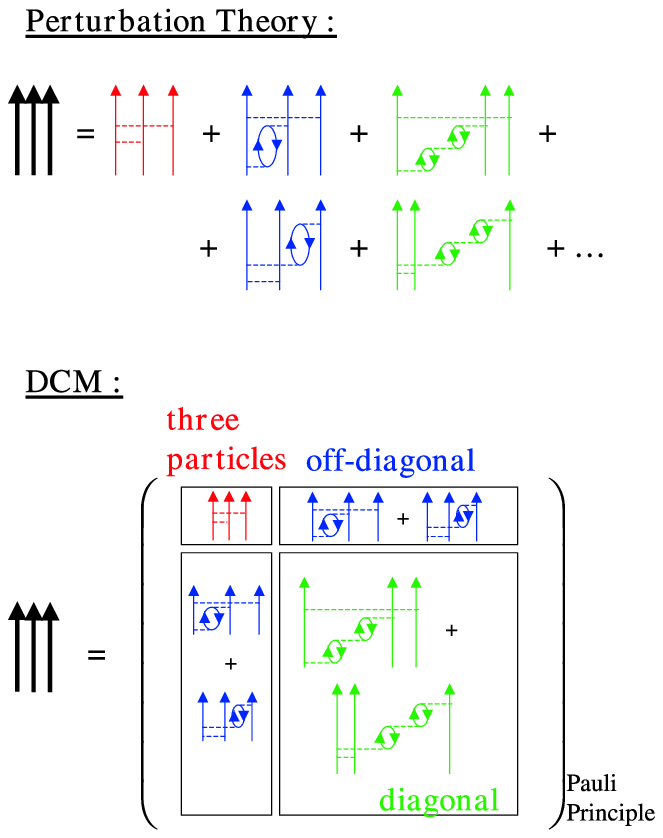}
\caption{\label{DP} Simbolic representation of {\it dressed}
three particles.}
\end{figure}

\begin{figure}
\epsfbox{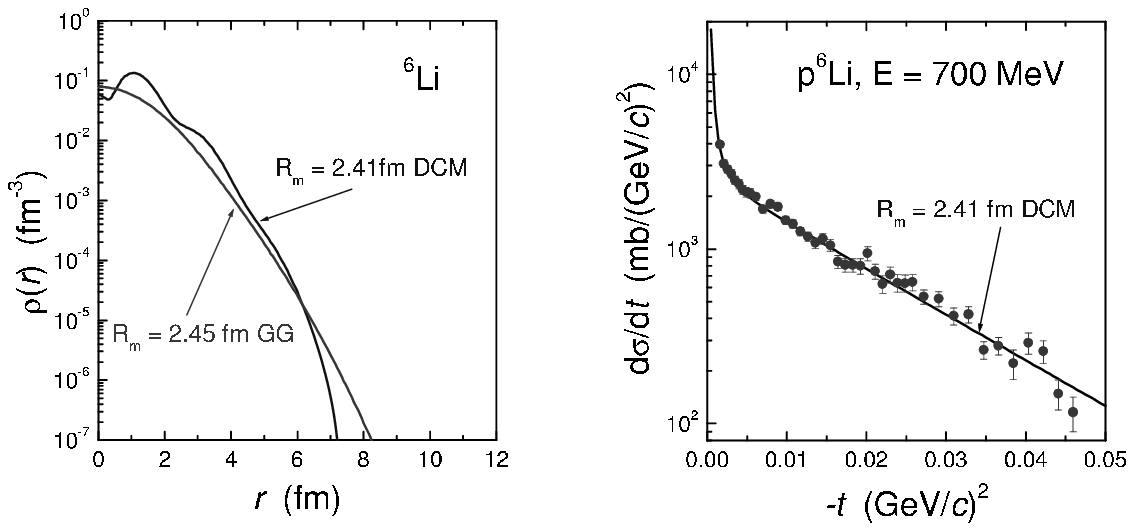} \vspace{-13.cm} \caption{\label{LIM1} Left:
Experimental and calculated distributions of $^{6}$Li; Right:
experimental and theoretical proton scattering cross sections.}
\end{figure}

\begin{figure}
\epsfbox{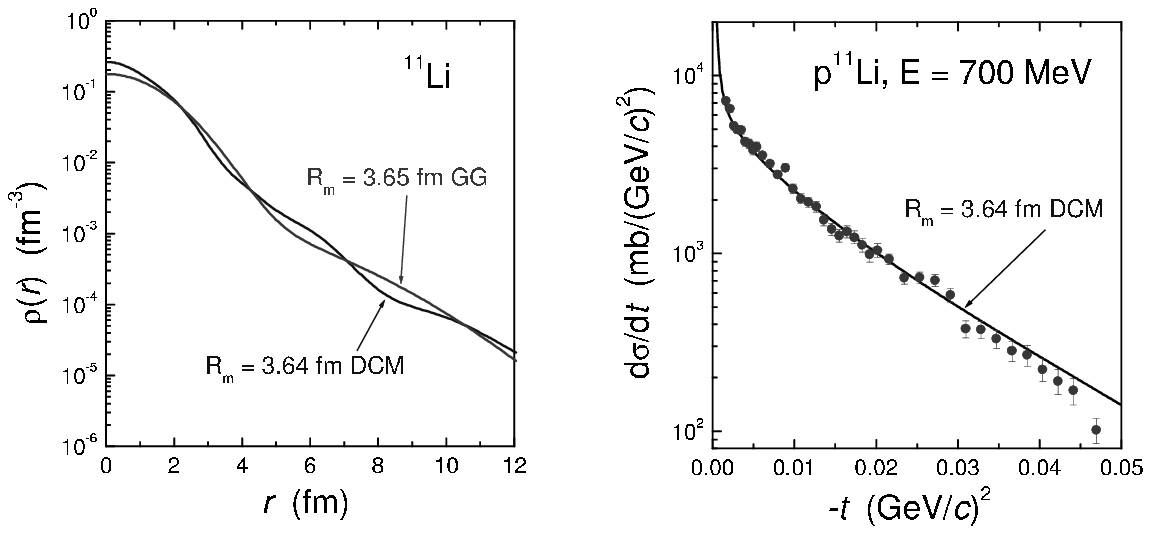} \vspace{-13.cm} \caption{\label{LIM} Left:
Experimental and calculated distributions of $^{11}$Li; Right:
experimental and theoretical proton scattering cross sections.}
\end{figure}

The transformation coefficients find also their applications in
the calculation of the matter distribution of halo nuclei~[20].
Here the calculation are performed for $^6$Li and $^{11}$Li
considering a large dimensional space and treating consistently
the interaction of the valence particles with the core
excitation. For $^{11}$Li, in order to reproduce the three-body
force we have performed calculations with three {\it dressed}
neutrons~[21], which are solutions of the  symbolic eigenvalue
equation given in Fig.~\ref{DP}. The results of the calculations
are given in Fig.~\ref{LIM1} and Fig.~\ref{LIM}. In
Fig.~\ref{LIM1}, we plot: (left) the experimental matter
distribution calculated in Ref.~[3] with the BDCM and the
experimental matter distribution; and (right) the {\it p}-$^6$Li
experimental scattering cross sections~[22] compared with the
results obtained in Ref.~[3] for the theoretical cross section.
In Fig.~\ref{LIM}, we plot: (left) the calculated~[3] and the
experimental matter distributions; (right) the calculated {\it
p}-$^{11}$Li scattering cross sections~[21] compared with the
experimental one [22]. For $^6$Li the calculated and the
experimental matter radii given in Fig.~\ref{LIM1} are in good
agreement with the experimental matter radius given in the figure.
The calculated matter radius of $^{11}$Li is 3.64 fm~[21] also in
good agreement with the experimental value of 3.62(19) fm~[22].
The BDCM and the DCM results obtained for the scattering cross
sections are in good agreement with the data of Ref.~[22]. For
$^{11}$Li the agreement is better for momentum transfer smaller
than .04 (GeV/c)$^2$, as for momentum transfer between 0.04-0.05
(GeV/c)$^2$. This is probably due to high-lying core states,
which are not included in the present DCM calculations, but are of
increasing importance at higher momentum transfer. The
consideration of these core states will increase drastically the
dimension of the CMWFs; in the calculations presented here
already about two hundred components were used. However, such
calculations presently under investigation become easily feasiable
with the use of cluster transformation coefficients.

\section{Proton polarization}
\begin{figure}
\epsfbox{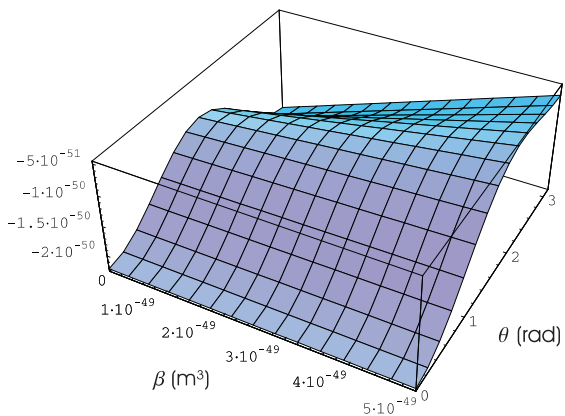} \caption{\label{SCCR} Scattering cross
sections of high energy photons as functions of scattering angle
and proton polarization parameter.}
\end{figure}

The scattering cross section of low-energy photons by a
structureless charged system is given by the Thomson formula
which derives the photon scattering cross section only in terms
of linear contributions which are proportional to the frequency
of the incoming photon. A second class of terms which are linear
in the frequency of the incident photon and take care of the
magnetic moment and spin of the target have also been included as
corrective factor to the Thomson scattering. Additionally to
these linear effects, however, the cross section get contributions
from terms proportional to the square of the photon frequency.
These terms, are the polarization terms of Ref.~[23] and modify
the cross section accordingly to the square bracket of the
following equation:
\begin{eqnarray}
\label{equ-ii-80} \fl {\frac{\delta \sigma}{\delta \Omega} =
(\frac{\delta \sigma}{\delta \Omega})_{point} + \left [ -r \left
(\frac{E_{sc}}{E_i} \right )^2 \frac{E_{sc}E_i}{(\hbar c)^2}
\left (\frac{\alpha + \beta}{2}(1+\cos\theta)^2+
\frac{\alpha-\beta}{2}(1-\cos\theta)^2 \right ) \right ] }
\end{eqnarray}
In Eq.~(\ref{equ-ii-80}), the first term $(\frac{\delta
\sigma}{\delta \Omega})_{point}$ give the scattering of photons
from a point-like hadron corrected with the size-spin
contributions, r is the classical radius of the proton, $E_i$ is
the incidental photon energy, $E_{sc}$ the photon scattering
energy, $\theta$ the scattering angle, $\alpha$ the electric- and
$\beta$ the magnetic-polarizability {\i.e.}: terms quadratic in
the photon frequency.
 In Fig.~\ref{SCCR}, we plot the polarization cross section using for $\alpha$
and $\beta$ the values calculated with the model independent
dispersion sum rule in Ref.~[24]:
\begin{eqnarray}
\label{equ-ii-81} \fl {(\alpha + \beta)=(14.2 \pm 0.03)\times
10^{-49} m^3}
\end{eqnarray}

These values are around those measured for the proton electric-
and magnetic-polarization in Ref. [25]. The plotted polarization
terms are ca. twelve order of magnitude smaller then the Compton
term. Under this consideration with the present parameters of the
PHELIX-laser in mind it is difficult to perform measurements to
obtain the proton polarizability within a better resolution.
However, additionally to the polarization effect produced by the
energy of the incoming photon field, an intrinsic polarization
effect~[26] may contribute to the scattering cross section. This
would result from effects analogous to the {\it boiling} of the
QED terms. The three quarks are exciting from the vacuum in a
nonlinear mechanism the $\langle q \overline{q}\rangle$
condensate. The three {\it dressed} quarks are also eigenvalues
of the equation symbolically represented in Fig.~\ref{DP}.
Calculations are in progress using for the quark the harmonic
oscillator base proposed in Ref.~[27]. The QCD boiling terms
should then give rise to a radial distribution with an halo shape
as calculated in Fig.~\ref{LIM} for $^{11}$Li. De facto the
intrinsic polarization effect increases the polarizability of the
proton and may produce a sizable effect for the present
parameters of PHELIX-laser.
 Therefore, considering this new source of polarization, experiments
can reveal unexpected results.
\section{Conclusion}
The linearization approximation and the cluster transformation
coefficients are important tools in solving nonlinear systems
such as those characterizing {\it n dressed} particles: the {\it
linearization ansatz} defines a collective hamiltonian that is
selfconsistently solvable for the model eigenvalues, while the
{\it transformation coefficients} provide easy computation for the
complex {\it n}-body matrix elements which are the input to the
collective eigenvalue equation. Within the linearization ansatz
and the transformation coefficients, new theoretical insight have
been obtained for the physics of interacting particles which
coexist with the excitations of the model vacuum.

\begin{table}
\caption{\label{table 1}Coefficients for the reduction $3p
\Rightarrow(pp)_{J_r}\otimes(p)_{J_s}$} \label{3p}
\begin{tabular}{rlr}

$V_{3,3}((3)J_1(3),T_1(1); (5,5)J_2(1),T_2(0)|\}$ &
$(5,5)J_r=1,T_r=1;
(3)J_s=3,T_s=1)$ &  -0.1362 \\
                                                  & $(5,5)J_r=3,T_r=1;
(3)J_s=3,T_s=1)$ &  -0.0422 \\
                                                  & $(5,3)J_r=1,T_r=1;
(5)J_s=5,T_s=1)$ &   0.8613 \\
                                                  & $(5,3)J_r=2,T_r=1;
(5)J_s=5,T_s=1)$ &  -0.4876 \\
\end{tabular}
\end{table}

\begin{table}
\caption{\label{table 2}Coefficients for the reduction $3p1h
\Rightarrow(ph)_{J_r}\otimes(pp)_{J_s}$} \label{3p1h1}
\begin{tabular}{rll}
 $T_{1,0}((3,5)J_1(2),T_1(0); (1,1)J_2(1),T_2(0)|\}$ & $(3,1)J_r=1,T_r=0;
(5,3)J_s=1,T_s=0)$ & 0.9326 \\
                                                     & $(3,1)J_r=2,T_r=0;
(5,3)J_s=2,T_s=0)$ & 0.3606 \\
\end{tabular}
\end{table}

\begin{table}
\caption{\label{table 3}Coefficients for the reduction $3p1h
\Rightarrow(pp)_{J_r}\otimes(ph)_{J_s}$} \label{3p1h2}
\begin{tabular}{rll}
 $V_{1,0}((3,5)J_1(2),T_1(0); (1,1)J_2(1),T_2(0)|\}$ & $(3,5)J_r=1,T_r=1;
(3,1)J_s=2,T_s=0)$ & 0.1397 \\
                                                     & $(3,5)J_r=3,T_r=1;
(3,1)J_s=2,T_s=1)$ & 0.9902 \\
\end{tabular}
\end{table}

\begin{table}
\caption{\label{table 4}Coefficients for the reduction $4p1h
\Rightarrow (3p)_{J_r}\otimes(ph)_{J_s}$} \label{4p1h1}
\begin{tabular}{ll}
 $V_{3,3}((1);(1,1)J_1(1),T_1(0); (5,1)J_2(2),T_2(1); J_3(1),T_3(1)| \}$ &  \\
 $(1,1,1)J_r=1,T_r=1; (5,1)J_s=1,T_s=1)$  & 0.9513 \\
 $(1,1,5)J_r=3,T_r=2; (1,1)J_s=1,T_s=1)$  & 0.3081 \\
\end{tabular}
\end{table}

\begin{table}
\caption{\label{table 5}Coefficients for the reduction $4p1h
\Rightarrow (2p1h)_{J_r}\otimes (pp)_{J_s} $} \label{4p1h2}
\begin{tabular}{ll}
 $W_{3,3}((1)(3,5)J_1(1),T_1(0); (5,1)J_2(2),T_2(1); J_3(1),T_3(1)| \}$ & \\
 $(1,1,1)J_r=1,T_r=1; (1,5)J_s=1,T_s=1)$   & 1.0     \\
\end{tabular}
\end{table}
\begin{table}[th]
\caption{Ground state hyperfine structure splittings for a point-nucleus ($%
E_{{\rm PN}}$), including Breit-Schawlow ($E_{{\rm BS}}$), DCM,
and QED corrections. The QED contributions include vacuum
polarization and self energy.  All values are in eV.}
\label{tab-DCM-QED}
\begin{tabular}{lcccccc}
& $^{185}$Re$^{74+}$ & $^{187}$Re$^{74+}$ & $^{203}$Tl$^{80+}$ &
$^{205}$Tl$^{80+}$ & $^{207}$Pb$^{81+}$ & $^{209}$Bi$^{82+}$ \\
\hline
$E_{\rm PN}$ & 3.0103 & 3.0411 & 3.0184 & 2.9890 & 1.3998 & 5.8395 \\
$E_{\rm BS}$ & 2.7976 & 2.8263 & 3.3073 & 3.3374 & 1.2528 & 5.1922 \\
$E_{\rm tot}^{{DCM}}$ & 2.7192 & 2.7449 & 3.2130 & 3.2770 & 1.2166 & 5.0832 \\
\hline $\Delta E_{\rm QED}$  & -0.0142 & -0.0143 & -0.0176 &
-0.0177 & -0.0067 & -0.0280
\\ \hline
$E_{\rm tot}$ & 2.7050 & 2.7306 & 3.1954 & 3.2213  &
1.2099 & 5.0552 \\
$E_{\rm exp}$ & 2.7187~(18) & 2.7449~(18) & 3.21351(25) &
3.24409(29) &
1.2159~(2) & 5.0841~(4) \\
Ref. & \hphantom{.}[11] & \hphantom{.} [11] & [12] & [12] &
\hphantom{.}[13] & \hphantom{.}[14]
\end{tabular}
\end{table}
\begin{table}
\caption{\label{table6}Comparison of different hfs calculations
in meV for Li-like bismuth with experimental HFS splitting}
\label{lilike}
\begin{tabular}{llll}
 Contribution (meV)    & {[17]}  & {[18]}    & {[16]} \\
 one-electron      &   958.50\,(5)    & 958.49        & 958.51 \\
 charge distr.     &  -113.8\,(3)     & -151.44         &  -113.61 \\
 mag. distr.       &   -13.9\,(2)     & -18.68         &  -14.1 \\
 total QED         &    -4.44         & -4.06          &  -4.81 \\
 e-e interaction   &    -29.45\,(4)   & 8.43           &  -3.4 \\
 boiling of QED vacuum          &     & 0.06           & -7.69 \\
 total theory      &    796.9\,(2)    & 792.8         & 783.9\,(3.0)  \\
 measurement [29] & & {820\,(26) meV}          &  \\
\end{tabular}
\end{table}

\References
\item[[1]  ] A.L. Fetter and J.D. Walecka, {\it Quantum Theory of Many-Particle
 Systems}, McGraw-Hill, New York (1971).
\item[[2]  ]  M. Tomaselli, Phys. Rev. {\bf C37}, 349 (1988);
            M. Tomaselli, Ann. Phys. (N.Y.) {\bf 205}, 362 (1991).
\item[[3]  ]  M. Tomaselli, L.C. Liu, T. K\"uhl {\it et al.}, submittet to PRC
(2002).
\item[[4]  ]  M. Tomaselli, Phys. Rev. {\bf C48}, 2290 (1993).
\item[[5]  ]  T. Erikson, K.F. Quader, G.E. Brown, and H.T. Fortune, Nucl. Phys.
 {\bf A465}, 123 (1987).
\item[[6]  ]  G.E. Brown, {\it Unified Theory of Nuclear Models}, North-Holland,
Amsterdam
 1964; A.M. Lane, {\it Nuclear Theory}, W.A. Benjamin Inc., New York (1964).
\item[[7]  ]  D.J. Rowe, {\it Nuclear Collective Motion}, Methmen and Co. Ltd.,
London (1970).
\item[[8]  ]  E.P. Wigner, {\it Group theory}, Academic Press, New York (1959).
\item[[9]  ]  G. Racah, CERN Report No. 61.8 (1961);
A. de-Shalit and I. Talmi,{\it Nuclear Shell Theory}, Academic
Press, New
 York (1963).
\item[[10]  ]   M. Tomaselli, S.M. Schneider, E. Kankeleit, and T. K\"uhl, Phys.
Rev.
 {\bf C51}, 2989 (1995);
 M. Tomaselli, T. K\"uhl, P. Seelig, C. Holbrow, and E. Kankeleit, Phys. Rev.
 {\bf C58}, 1524 (1998);
 M. Tomaselli, T. K\"uhl, W. N\"ortersh\"auser, S. Borneis, A. Dax,
 D. Marx, H. Wang, and S. Fritzsche, Phys. Rev. {\bf A65}, 022502 (2002).
\item[[11]  ]  J. R. Crespo L\'{o}pez-Urrutia, P. Beiersdorfer, K.
Widmann, B. B. Birkett, A.-M. M\aa rtensson-Pendrill and M. G. H.
Gustavsson, Phys. Rev. A {\bf 57}, 879 (1998).
\item[[12]  ] P. Beiersdorfer et al., Phys. Rev. A {\bf 64}, 032506 (2001).
\item[[13]  ]  I. Klaft {\it et al.}, Phys. Rev. Lett. {\bf 73}, 2425
(1993).
\item[[14]  ]  P. Seelig {\it et al.,} Phys. Rev. Lett. {\bf 81}, 4824
(1998).
\item[[15]  ]  V.M. Shabaev, M. Tomaselli, T. K\"uhl, A.N. Artemyev, and V.A.
Yerokhin, Phys. Rev. A {\bf 56}, 252 (1997); S. M. Schneider, W.
Greiner, G. Soff, Phys.
Rev. A {\bf 50}%
, 118 (1994); H. Person, S. M. Schneider, G. Soff, W. Greiner,
Phys. Rev. Lett. {\bf 76}, 1433 (1996); P. Sunergren, H. Persson,
S. Salomonson, S. M. Schneider, I. Lindgren, and G. Soff, Phys.
Rev. A {\bf 58}, 1055 (1998).
\item{[16]  }  M. Tomaselli, T. K\"uhl, W. N\"ortersh\"auser, G. Ewald, R.
Sanchez,
             S. Fritzsche, and S.G. Karshenboim, Can. J. Phys., 2002 to be
published.
\item[[17]  ]  V. M. Shabaev et al., Phys. Rev. A {\bf 57}, 149 (1998).
\item[[18]  ]  S. Bouchard and P. Indelicato, Eur. Phys. J. D {\bf 8}, 59
(2000).
\item[[19]  ]  P. Beiersdorfer et al., Phys. Rev. Lett. {\bf 80}, 3082 (1998).
\item[[20]  ]  I. Tanihata {\it et al.}, Nucl. Phys. {\bf A488}, 113 (1988).
\item[[21]  ]  M. Tomaselli, T. K\"uhl, P. Egelhof, W. N\"ortersh\"auser,
 C. Kozhuharov, A. Dax, H. Wang, S.R. Neumaier, D. Marx, H.-J. Kluge
 I. Tanihata, S. Fritzsche, and M. Mutterer, Proc. Conf. Nuclear Physics at
Border Lines
 (NPBL), Lipari (Messina), Italy, 2001, Edited G. Fazio, G. Giardina, F.
Hanappe,
  G. Imme' and N. Rowley, World Scientific, ISBN 981-02-4778-8, 336 (2002);
 M. Tomaselli {\it et al.}, Proc. Conf. Dynamical Aspects of
 Nuclear Fission (DANF), $\check{C}$ast$\acute{a}$-Papiernicka (2001), in press.
\item[[22]  ]  A.V. Dobrowolsky {\it et al.}, Proc. Conf. Nuclear Physics at
Border Lines
       (NPB), Lipari (Messina), Italy, 2001, Edited G. Fazio, G.
       Giardina, F. Hanappe, G. Imme' and N. Rowley, World Scientific, ISBN
981-02-4778-8, 336 (2002).
\item[[23]  ]  V.A. Petrun'kin, Soviet Physics {\bf JETP 13}, 808 (1961).
\item[[24]  ]  F.J. Federspiel et al., Phys. Rev. Lett. {\bf67}, 1511 (1991).
\item[[25]  ]  V. Olmos de Leon et al., Eur. Phys. J. {\bf A10}, 207 (2001).
\item[[26]  ]  M. Tomaselli, L. C. Liu, T. K\"uhl, D. Ursescu,
in preparation.
\item[[27]  ]  P. Geiger and N. Isgur, Phys. Rev. {\bf D55}, 299 (1997).

\endrefs

\end{document}